\documentclass[aip,  superscriptaddress,  reprint]{revtex4-1}

\usepackage{amsmath}
\usepackage{amssymb}

\usepackage{graphicx}
\usepackage{dcolumn}
\usepackage{bm}
\usepackage{hyperref}
\usepackage[absolute]{textpos}

\begin{document}

\title{Multistability, local pattern formation, and global collective firing in a small-world network of non-leaky integrate-and-fire neurons}
\author{Alexander \surname{Rothkegel}}
\email{alexander@rothkegel.de}
\affiliation{Department of Epileptology, University of Bonn, Sigmund-Freud-Str. 25, 53105 Bonn, Germany}
\affiliation{Helmholtz-Institute for Radiation and Nuclear Physics, University of Bonn, Nussallee 14-16, 53115 Bonn, Germany}

\author{Klaus \surname{Lehnertz}}
\email{klaus.lehnertz@ukb.uni-bonn.de}
\affiliation{Department of Epileptology, University of Bonn, Sigmund-Freud-Str. 25, 53105 Bonn, Germany}
\affiliation{Helmholtz-Institute for Radiation and Nuclear Physics, University of Bonn, Nussallee 14-16, 53115 Bonn, Germany}
\affiliation{Interdisciplinary Center for Complex Systems, University of Bonn, R{\"o}merstr. 164, 53117 Bonn, Germany} %
\date{\today}

\begin{abstract}
We investigate numerically the collective dynamical behavior of pulse-coupled non-leaky integrate-and-fire-neurons that are arranged on a two-dimensional small-world network. To ensure ongoing activity, we impose a probability for spontaneous firing for each neuron. We study network dynamics evolving from different sets of initial conditions in dependence on coupling strength and rewiring probability. Beside a homogeneous equilibrium state for low coupling strength, we observe different local patterns including cyclic waves, spiral waves, and turbulent-like patterns, which -- depending on network parameters -- interfere with the global collective firing of the neurons. We attribute the various network dynamics to distinct regimes in the parameter space. For the same network parameters different network dynamics can be observed depending on the set of initial conditions only. Such a multistable behavior and the interplay between local pattern formation and global collective firing may be attributable to the spatiotemporal dynamics of biological networks.
\end{abstract}
\maketitle

\begin{textblock*}{20cm}(0.5cm,26cm)
Copyright (2009) American Institute of Physics. This article may be downloaded for personal use only. Any other use requires prior permission of the author and the American Institute of Physics. The following article appeared in Chaos 19, 015109 (2009) and may be found at \url{http://link.aip.org/link/?cha/19/015109}.
\end{textblock*}

\textbf{
Pattern formation in neural networks plays a prominent role in understanding physiological and pathophysiological aspects of mammalian hearts and brains. In the case of the heart, normal functioning is determined by collective oscillations of the contributing cardiac cells, while ventricular and atrial fibrillation is related to the emergence of spiral wave patterns. People that suffer from migraine or are influenced by certain drugs report
on spiral wave patterns during visual hallucination, and these patterns can even be observed in the neuronal activity of the mammalian cortex. In addition, a multitude of brain disorders such as epilepsy, schizophrenia, autism, migraine, and Alzheimer's and Parkinson's disease are associated with abnormal collective firing emerging from neural tissue. We here observe the cooccurrence of local wave patterns and global collective firing 
in a two-dimensional small-world network composed of simple model neurons.  Our observations might be of relevance to gain deeper insights into how the spatiotemporal dynamics of brain disorders (e.g. epileptic seizures) depends on both the dynamic properties of neural elements and the topology of synaptic wiring.}

\section{Introduction}

A regular lattice with the local dynamics of excitable elements is called excitable medium \cite{Meron1992,Cross1993}. Systems that have been modeled as excitable media are ubiquitous in nature, ranging from isothermal chemical reactions \cite{Zaikin1970} via disease spreading among a population of living organisms \cite{Vannucchi2004} to the mammalian heart \cite{Gray1998,Kanakov2007} and brain systems \cite{Hemmen2004,Huang2004a,Dahlem2004,Dahlem2008}.
For many natural systems, however, the consideration of regular lattices may not yield an adequate description given that distant elements may interact.
Watts and Strogatz introduced model networks that take into account both local and long-range interactions \cite{Watts1998}. The authors start with a regular lattice and rewire some of the connections to random positions yielding a small-world configuration. Varying the fraction of rewired connections allows to interpolate continuously between regular lattices and random networks. This scheme has inspired many studies onto how the dynamics of coupled, complex networks changes along this interpolation \cite{Newman2003,Boccaletti2006a,Arenas2008,Dorogovtsev2008}. 

Small-world models have recently been shown to provide a useful framework that may help to improve our understanding of structure and function of human brain systems \cite{Bassett2006b}. Apart from the underlying network topology the dynamical properties of network elements can be regarded crucial for the collective dynamical behavior. Many neuron models have been proposed with varying numerical complexity \cite{Rabinovich2006,Izhikevich2007}.  Especially for detailed models and large networks, feasibility is easily lost. However, qualitative observations are often transferable between different neuron models, and even simple models like the integrate-and-fire (IF) neuron \cite{Burkitt2006a,Burkitt2006b} are powerful tools in understanding the information processing capabilities of real neurons. Despite their simplicity, analytical results for the global dynamical behavior of IF neuron networks are limited to special cases, mostly considering homogeneous configurations like all-to-all coupling \cite{Mirollo1990}, random networks \cite{Golomb2000}, or population models \cite{Sirovich2006}.  For lattices, wave propagation and spiral waves can be observed \cite{Horn1997}, and it is possible to perform a continuum limit, describing the medium in the form of a partial differential equation \cite{Coombes2005}. For the small-world regime, however, it is not clear how such a description can be achieved thus rendering numerical simulations inevitable (see e.g. \cite{Masuda2004}).  

To ensure ongoing activity in the medium, neurons are usually excited via some noise input. The existence of an optimal noise level for wave phenomena is called coherence resonance and has been extensively studied, also more recently for small-world media \cite{He2002,Perc2007,Sun2008}.  In two or more dimensions, the formation of self-sustaining activity such as spiral waves and irregular turbulent-like patterns is possible, even in the absence of an external input. In small-world media self-sustaining activity is also possible for one spatial dimension if the chosen setup allows for a balance between wave propagation and creation due to long-range excitations \cite{Roxin2004,Riecke2007}. More recently, it was reported that waves in a one-dimensional small-world network of phase oscillators can prevent synchronous motion for sufficiently large coupling strength and few random connections \cite{Park2007}. Similarly, a sharp transition between wave-dominated behavior for few random connections and collective periodic behavior for many random connections was observed for a two-dimensional coupled map lattice of excitable elements \cite{Sinha2007}.

Our work focuses on such transitions. Instead of strict nearest-neighbor coupling, however, we here consider the case of a radius of influence for nodes on the regular lattice which underlie our small-world networks. This choice is motivated by neuroanatomy  (cf. \cite{Braitenberg1991,Murre1995}) and leads to a complicated dependence of the dynamics (e.g. the propagation speed of waves (c.f. \cite{Bressloff2000})) on the coupling strength. For each node we here consider non-leaky IF neurons, and we impose a probability for spontaneous firing to each neuron which can be thought to arise from incoming synaptic excitation from outside the network. We study the occurring patterns and their influence on the transition between wave-dominated firing and global collective firing of neurons. We distinguish between self-sustaining patterns like spiral waves, which emerge due to local excitations, and cyclic waves, which emerge due to long-range excitations. 

This article is organized as follows. In Sec. \ref{sec:methods} we give a detailed description of our dynamical system and the chosen observables. In Secs. \ref{sec:result_lattice} and \ref{sec:result_random} we discuss the behavior for regular lattices and random networks separately  before continuing with the investigation of small-world networks in Sec. \ref{sec:result_smallworld}. We finally draw our conclusions in Sec. \ref{sec:conclusion}.

\section{Methods}
\label{sec:methods}

We here consider a two-dimensional regular lattice of $N = 300 \times 300$ identical, non-leaky IF neurons. Two different neurons are said to be connected if their Euclidean distance is smaller or equal than the radius of influence $R$.
In the following we present our findings for cyclic boundaries and note that we observed similar dynamical behavior for open boundaries. We also note that we obtained qualitatively similar findings for smaller network sizes ($100 \times 100$ and $200 \times 200$ neurons) and thus expect that our findings carry over to larger network sizes. Starting from this configuration, every directed connection is removed with probability $\rho \in [0,1]$  and a connection between two randomly chosen, unconnected neurons $n_1 \neq n_2$ is introduced. With this rewiring scheme the mean degree $d$ is independent on $\rho$. For two neurons $n_1, n_2$ that are connected via a synapse from $n_1$ to $n_2$, we write $n_1 \triangleleft n_2$. The dynamical state of each neuron in the network at time $t$ is fully determined by its membrane potential $x_n(t)$. Negative values of $x_n$ signify refractoriness; neurons for which $x_n \geq \vartheta$ fire and increase the membrane potential of all neurons $n'$ with $n \triangleleft n'$ by the global coupling strength $c$. The number of time steps during which neurons remain refractory after firing will be denoted by $\tau$. To every neuron $n$ we associate a bimodal random variable $\eta_n(t)$, which takes a value of $1$ with probability $p_{\mbox{s}}$ and $0$ otherwise. $\eta_n(t)$ determines the times at which neuron $n$ fires spontaneously. We define for every neuron $n$ the number of firing neurons which are connected by incoming synapses as $f_n(t) = |\{n' | x_{n'}(t) > \vartheta, n' \triangleleft n \} | $ . The dynamics of neuron $n$ in discrete time $t$ can now be described as:
\[
 x_n(t+1) = \begin{cases}
		x_n(t) + 1 & x_n(t) < 0, \\
		x_n(t) +   \eta_n(t)\vartheta+ c f_{n}(t)& 0 \leq x_n(t) <\vartheta, \\
		- \tau & x_n(t) \geq\vartheta.
            \end{cases}
\]

After choosing a membrane potential $x_n(0)$ for every neuron as its initial condition, the coupled dynamical system is iterated for $T$ time steps. Mostly, we will consider here the evolution of a homogeneous state with identical membrane potentials for all neurons, i.e. $x_n(0) = 0$. 

\begin{figure}
\includegraphics{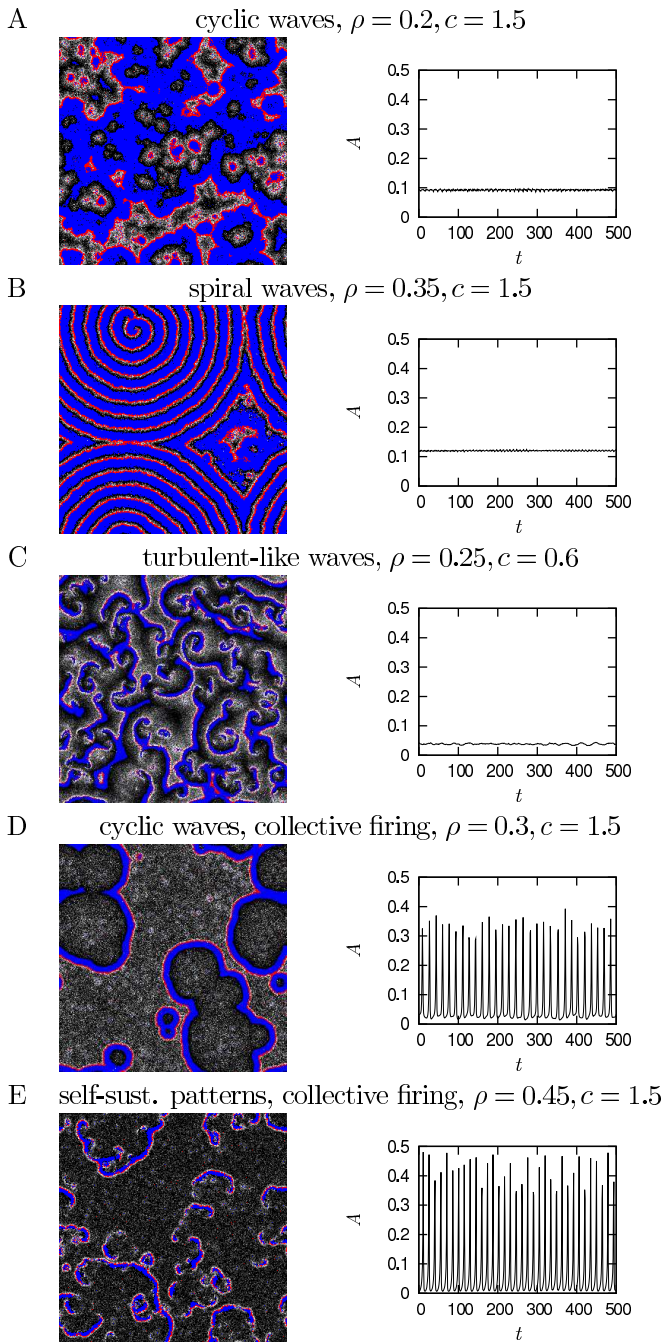}
\caption{\scriptsize{(Color online) Exemplary snapshots of the spatial distribution of membrane potentials (left) together with the corresponding network activity $A(t)$ (right) for small-world networks with different rewiring probabilities $\rho$ and coupling strengths $c$. Initial conditions: $x_n(0) = 0$. In the snapshots neurons are indicated as points on the coordinates of the underlying lattice. Red points correspond to firing neurons, blue points indicate refractory neurons, and gray points denote charging neurons with lightness encoding the membrane potential. A: Several foci emit cyclic waves at a specific temporal order.  Colliding waves annihilate because of the refractoriness of the medium. The resulting network activity is of small variance. B: Spiral waves dominate the dynamics. They lead to a network activity $A(t)$ with small variance but the temporal average is slightly increased as compared to cyclic waves.  C: The dynamics shows irregular turbulent-like patterns leading to a network activity with small variance. D: The majority of the neurons charge and fire collectively. Randomly, cyclic waves are created. The corresponding network activity shows a periodic behavior. E: Mostly collective firing of neurons with some self-sustaining patterns and periodic network activity.  \label{latticeSnapshotsSmallworld}}}
\end{figure}

As observable we use the fraction of firing neurons per time step $A(t)$, which will be denoted as network activity. In Fig. \ref{latticeSnapshotsSmallworld} we present exemplary snapshots from the temporal evolution of the spatial distribution of membrane potentials along with the network activity for different dynamical scenarios. 

As only the ratio between $c$ and $\vartheta$ influences the dynamics, we set, for the sake of simplicity, $\vartheta  = 10$ in all simulations. Note, that large values of $R$ and $\tau$ increase the size of all wave phenomena. Therefore, a trade-off between discretization and finite-size-effect has to be made and we chose $\tau = 5$ and $R = \sqrt{10}$. For this choice every neuron in the middle of the lattice is connected to 36 neurons. The probability for spontaneous firing was chosen to be small in the sense that the influence on the dynamics is mainly to ensure that the activity in the medium does not die out. We set $p_{\mbox{s}} = 0.001$. The remaining two free system parameter, namely the coupling strength $c$ and the rewiring probability $\rho$, were varied in our simulations to estimate their influence on the dynamics. We will refer to a small-world network with the local dynamics of an excitable element as small-world medium.

In parts D and E of Figs \ref{latticeSnapshotsSmallworld} a large fraction of neurons charges and fires collectively, which leads to alternating periods of low and high network activity.  We will connote such an oscillatory behavior with \emph{global synchrony} and distinguish it from a behavior as in parts A, B, and C of Fig. \ref{latticeSnapshotsSmallworld}, where the network activity exhibits only minor fluctuations over time. In order to classify the network dynamics, we define the following order parameter which takes large values for global synchrony:
\[
 	r =  \max ( A(t) | 0 \leq t \leq T )- \min ( A(t) | 0 \leq t \leq T).
\]
For large observation times $T$ and stationary dynamics, $r$ converges to the dynamical range of $A$. Note, that $r$ does not allow to differentiate between random and periodic network activities. However, the observed activities display large dynamical ranges of $A$ always combined with periodicity, which allows us to use this simple ansatz to detect collective firing of neurons.
To account for transients, we ignore the first $2000$ time steps of observation. We mention though that special care has to be taken since the observed system dynamics may change for certain parameter settings even after long periods of stationarity. 

Note that the characteristic time scale of the system is determined by different mechanisms for cyclic waves and for spiral wave or turbulent-like patterns. For cyclic waves the formation of a wavefront is caused by non-local inputs from spontaneous firing and long-range connections: For spiral waves and turbulent-like patterns the formation is caused by local connections and is in a wide parameter range nearly independent on $p_{\rm{s}}$ and $\rho$. Therefore, we expect different wave densities for both behaviors. Due to the ability of spiral waves and turbulent-like patterns to sustain even without external input ($p_{\rm{s}} = 0$), we will denote both of them as self-sustaining patterns. To distinguish between cyclic waves and self-sustaining patterns we use the mean firing rate $m$: 
\[
 	m = \frac{1}{T} \sum\limits_{t = 0}^T A(t).
\]

\section{Results}
\subsection{Lattices}
\label{sec:result_lattice}

For lattices, the dynamics is governed by local pattern formation known from excitable media. We observe random firing, more complex turbulent-like patterns as well as cyclic or spiral waves depending on the coupling strength $c$ and on the chosen set of initial conditions. Note that wave propagation in the medium requires a minimal coupling strength, which can be estimated for a wave traveling in either vertical or horizontal direction through an unexcited medium. Given our choice of the radius of influence $R = \sqrt{10}$, every neuron is connected to 15 neurons from the three rows below, to 15 neurons from the three rows above, and to 6 neurons in the row of the considered neuron. This leads to a minimal coupling strength of $\frac{\vartheta}{15} = 0.67$. For a medium that is already somewhat excited, wave propagation may be possible below this threshold. 

\begin{figure}
\includegraphics[width=0.85\columnwidth]{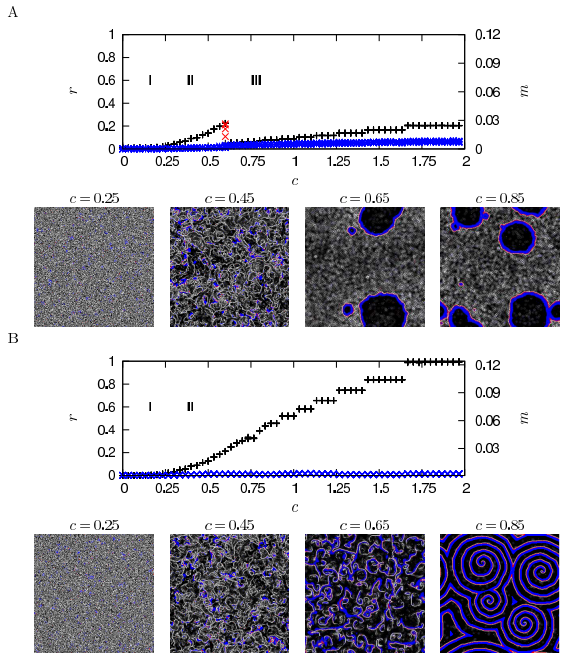}
\caption{\scriptsize{(Color online) A (upper part): Order parameter $r$ (blue) and mean firing rate $m$ (black) dependent on coupling strength $c$ for a regular lattice. Initial conditions: $x_n(0) = 0$. Red  symbols indicate values of $m$ and $r$ that originate from changes of the dynamical behavior during the observation time. The systems were observed for 8000 time steps with 20 realizations for each coupling strength. A (lower part): Snapshots of the spatial distribution of membrane potentials for different coupling strength (color coding as in Fig \ref{latticeSnapshotsSmallworld}). B: Same as in A but for open wave endings as initial conditions as described in Fig. \ref{openWaveEnding}. For $ c < 0.28$ the dynamics for both sets of initial conditions is dominated by spontaneous firing and we observe no pattern (regime I). For $0.28 < c < 0.6$ both sets of initial conditions lead to self-sustaining patterns (regime II). For $c > 0.6$, the initial conditions $x_n(0) = 0$ lead to cyclic waves (regime III), while open wave endings lead again to self-sustaining patterns (turbulent-like patterns and spiral waves). Note that because of the discrete nature of the observed dynamical system, the dynamics does not depend continuously on the coupling strength; it only changes when crossing a fraction of the threshold potential, which is reflected by stepwise constant firing rates $m$.} \label{meanOrderLattice}}
\end{figure}

With the set of initial conditions $x_n(0) =0$ for all neurons, the mean firing rate $m$ and the order parameter $r$ depend discontinuously on the coupling strength (cf. Fig. \ref{meanOrderLattice}A), and the medium shows different dynamical behaviors. Starting at $c=0$ the dynamics is dominated by spontaneous firing of neurons.  Here, the membrane potentials of the neurons in the charging state (with $0 < x_n(t) < \vartheta$) are distributed around their mean value $\bar{x}(t)$, which we denote as mean excitation in the network. As a spontaneously firing neuron reenters the charging state with $x_n (t)=0$ after the refractory period, the influence of the firing on the mean excitation amounts to $\Delta = 1/N(-\bar{x}(t) + d c)$. 
The mean excitation $\bar{x}(t)$ saturates for $\Delta = 0$ when charged by spontaneous firing only. From this condition the saturation potential 
$\tilde{x}$ can be estimated as $\tilde{x} = dc$. If $\tilde{x} < \vartheta$ or $c \leq \frac{\vartheta}{d} = \frac{10}{36}=0.28$ the charging saturates before the majority of neurons reach their threshold potential. Thus the dynamics is characterized as a homogeneous equilibrium without any observable pattern (regime I). For $0.28<c<0.6$ the dynamics gradually changes to turbulent-like patterns (regime II). Here, the medium is excited in a complicated way, and wave propagation is partially possible. The patterns preserve themselves because waves die out and leave the medium somewhat excited or because wave propagation is so slow that the neurons in the tail of the wave recover from their refractory period and get excited again. This dynamical behavior relies on $R$ and $\tau$ being of similar order of magnitude and can probably only be observed for $R>1$. Note that for $c > 0.47$ these turbulent-like patterns sustain even in the absence of external input ($p_{\rm{s}} = 0$). For $c > 0.6$, turbulent-like patterns do not appear anymore. Instead the medium is charged homogeneously (by the spontaneous firing) until at regions cyclic waves appear clearing large parts of the medium again from excitement (regime III). Given our choice of parameters the charging of the medium is slow as compared to the formation of turbulent-like patterns, which is reflected by a decreased mean firing rate $m$.

\begin{figure}
\includegraphics[width = \columnwidth]{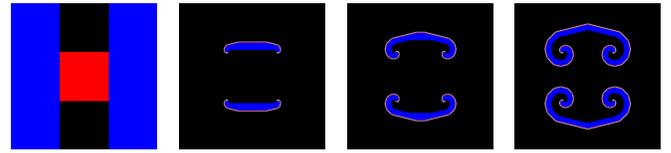}
\caption{\scriptsize{(Color online) Consecutive snapshots of the spatial distribution of membrane potentials for a regular lattice with coupling strength $c = 1.0$ and without spontaneous firing. The initial conditions are defined as follows: $x_n(0) = - \tau$ for the neurons in the left and the right third of the lattice, $x_n(0) = 0$ for neurons in the upper and lower part of the middle third, and $x_n(0) = \vartheta$ for the remaining neurons (color coding as in Fig. \ref{latticeSnapshotsSmallworld}). \label{openWaveEnding}}}
\end{figure}

In contrast, when starting from a set of initial conditions as depicted in Fig. \ref{openWaveEnding} the mean firing rate $m$ increases monotonously with the coupling strength $c$ and self-sustaining patterns can still be observed for $c > 0.6$ (cf. Fig \ref{meanOrderLattice}B). 
Here, the dynamics of the lattice is characterized by four open endings of two wavefronts that bend and create turbulent-like patterns. For larger coupling strengths the patterns become more regular until four spiral wave foci remain for coupling strengths $c > 0.8$.

For both sets of initial conditions the order parameter $r$ takes on small values only as we do not observe global oscillations on our lattices.

\subsection{Random networks}
\label{sec:result_random}
We generated random networks by connecting every pair of nodes with a fixed connection probability $\rho$. Although this leads to slightly varying total number of connections per realization, the random networks will be indexed here by their expected mean number of connections per neuron $d = \rho N$ (i.e., the mean degree). We observe two regimes that do not depend on the choice of the initial conditions. For small coupling strengths $c$ the network activity $A(t)$ is of small variance (regime a), and for large $c$ it shows a periodic behavior (regime b). Both dynamics are separated in parameter space by a critical coupling strength, which will be denoted by $c_c$.

\begin{figure}
\includegraphics{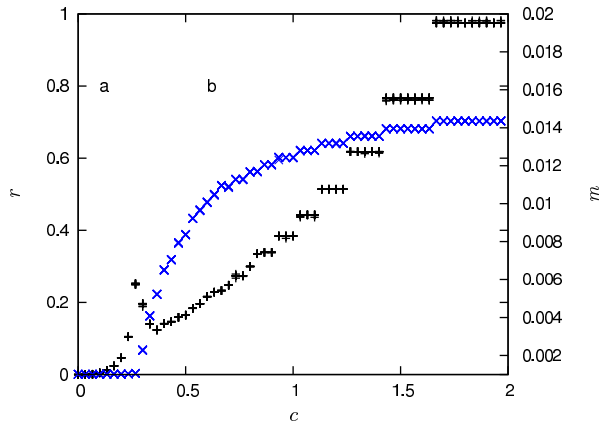}
\caption{\scriptsize{(Color online) Order parameter $r$ (blue) and mean firing rate $m$ (black) dependent on the coupling strength $c$ for a random network with mean degree $d = 36$. For $c > 0.28$ network activity $A(t)$ shows periodic behavior (regime b), while for smaller coupling strengths the network activity shows only minor fluctuations (regime a). We observe no dynamical changes during the observation time. \label{kuramotoKurve}}}
\end{figure}

The dependence of the order parameter $r$ on the coupling strength $c$ shows the typical behavior known from other synchronization phenomena \cite{Strogatz2000}. We present this dependence in Fig. \ref{kuramotoKurve} for $d=36$, which is the same mean number of connections as studied with regular lattices. The mean firing rate $m$ takes on similar values as observed for cyclic waves on a lattice. Interestingly, $m$ exhibits a local maximum at $c_c$. 

\begin{figure}
\includegraphics{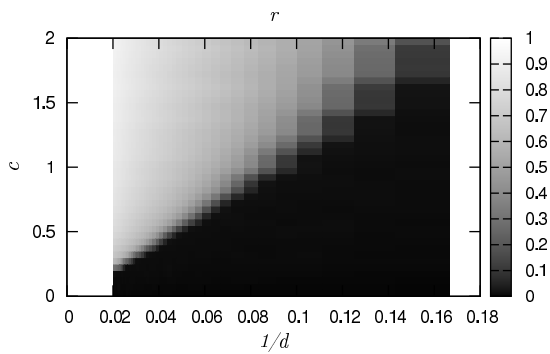}
\caption{\scriptsize{Order parameter $r$ dependent on the coupling strength $c$ and inverse mean degree $1/d$ for a network of 90000 randomly coupled neurons. The system was observed for 8000 time steps. \label{randomNetworkOrderParameter}}}
\end{figure}

From the dependence of the order parameter $r$ on the mean degree $d$ and on the coupling strength $c$ (cf. Fig. \ref{randomNetworkOrderParameter}), we observe that $c_c$ can be described with the parameters of the system in a simple way: $d c_c = \vartheta$. As with regular lattices spontaneous firing charges the mean excitation in the network $\bar{x}(t)$ only until $\tilde{x} = d c$. If the charging saturates below the firing threshold (i.e., for $dc < \vartheta$) neurons do not fire collectively. Note that as considered in \cite{Sirovich2006}, a probability for an excitation instead of for firing, would lead to an effective charging of the medium independently of $\bar{x}(t)$, and global oscillations of the network activity could be observed for $c < \vartheta/d$. The critical coupling $c_c$ also marks the threshold at which the network becomes non-dissipative in the sense that each neuron distributes more excitation than was needed to make it fire; in the case of IF neurons in continuous time and without refractoriness, coupling strengths $c > c_c$ would lead to divergent behavior.

\subsection{Small-world networks}
\label{sec:result_smallworld}

Before presenting our findings for the transition between regular lattices and random networks we briefly recall the main findings for the limiting cases. On a regular lattice we observe -- for homogeneous initial conditions -- different local patterns depending on the coupling strength $c$. For small $c$ we observe random firing (regime I), for intermediate $c$ self-sustaining patterns (regime II) and for large $c$ cyclic waves (regime III). The transition between regimes II and III can be assessed by a discontinuity in the dependence of the mean firing rate $m$ on $c$.  

For random networks we observe -- independent of the initial conditions -- a smooth transition between constant (regime a) and periodic network activities (regime b), as assessed by the order parameter $r$. The different dynamical behaviors can be expected to be carried over into the small-world regime and thus allow one to investigate the interplay between local patterns and global oscillations.

\begin{figure}
\includegraphics{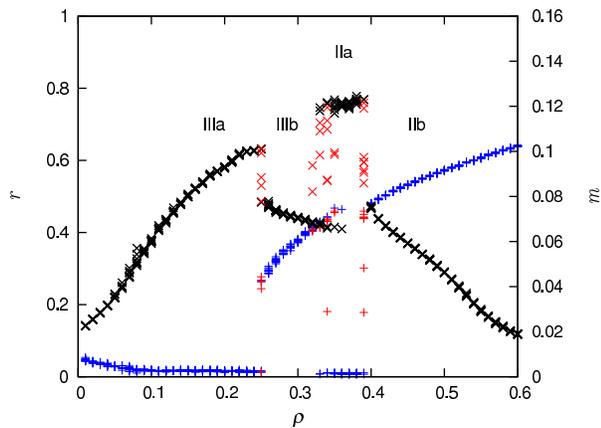}
\caption{\scriptsize{(Color online) Order parameter $r$ (blue) and mean firing rate $m$ (black) dependent on the rewiring probability $\rho$ for a small-world medium with coupling strength $c = 1.5$. Red  symbols indicate values of $m$ and $r$ that originate from dynamical changes during the observation time. Initial conditions: $x_n(0) = 0$ for all neurons. The system was observed for 8000 time steps, and 10 realizations with different random seeds for rewiring and spontaneous firing were simulated for each $\rho$. Four different regimes can be observed. Regime IIIa: no global oscillations, cyclic waves (cf. Fig. \ref{latticeSnapshotsSmallworld}A). Regime IIIb: global oscillations, cyclic waves (cf. Fig. \ref{latticeSnapshotsSmallworld}D). Regime IIa: no global oscillations, self-sustaining patterns (cf. Fig. \ref{latticeSnapshotsSmallworld}B). Regime IIb: global oscillations, self-sustaining patterns (cf. Fig. \ref{latticeSnapshotsSmallworld}E).\label{meanOrderCoupling1.5} }}
\end{figure}

First, we discuss the temporal evolution of the set of homogeneous initial conditions ($x_n(0) = 0$ for all neurons). In Fig. \ref{meanOrderCoupling1.5} we present the dependence of the mean firing rate $m$ and of the order parameter $r$ on the rewiring probability $\rho$ for a fixed coupling strength $c = 1.5$. We can separate four different regimes. Transitions between regimes can be assessed as discontinuities in $m$ and $r$. For each of the regimes, a snapshot of the spatial distribution of membrane potentials is presented in Fig. \ref{latticeSnapshotsSmallworld}. 
The dynamical behavior in these regimes differs in the local patterns (as observed in II and III on regular lattices) and in whether the medium exhibits global oscillations (as observed in a and b on random networks). We thus denote these regimes as IIa, IIb, IIIa, and IIIb. 

For small $\rho$ we observe no global oscillations. This is due to the influence of cyclic waves and could be explained by the following considerations. A wave will clear any excitement it passes on the medium. If the characteristic time $t_{\rm{w}}$ for a region of the medium between the passing of two waves is smaller than the time $t_{\rm{o}}$ that is needed to charge this region due to rewired connections and spontaneous firing, then the dynamics will evolve to a wave-dominated state with constant activity. With more rewired connections $t_{\rm{o}}$ decreases. Therefore, $t_{\rm{w}} = t_{\rm{o}}$ for some $\rho$ and the global oscillations become stable (transition between regime IIIa and IIIb). 

With more rewired connections also the number of local connections is diminished. At a certain rewiring probability the patterns change from cyclic waves to self-sustaining patterns. Especially after the global firing, waves get so slow that the neurons in the tail recover from their refractory period and get excited again. This excitation eventually forms a second wave, which collides with the refractory tail of the first wave creating spiral wave foci. As self-sustaining patterns lead to larger wave densities, $t_{\rm{w}}$ is diminished abruptly and the dynamics gets dominated by waves again (transition between regime IIIb and IIa).  

A wave in a medium exhibiting global oscillations will be {\em moved backward} by each oscillation for a distance it moves in the refractory period. This is because refractory neurons in the tail of a wave will not participate in the global collective firing and subsequently, the wave will reemerge behind this tail. Additionally, the propagation speed of a wave depends on the excitation of the medium. Therefore, speed will increase on the way to collective firing or a wave will only move part of the time when the membrane potentials are already charged.  At a certain rewiring probability the waves get so slow that they practically do not move anymore.  For this rewiring probability global oscillations become stable again (transition between regime IIa and IIb). Although some wave phenomena remain in regime IIb, their portion diminishes rapidly with increasing $\rho$.

\begin{figure}
\includegraphics{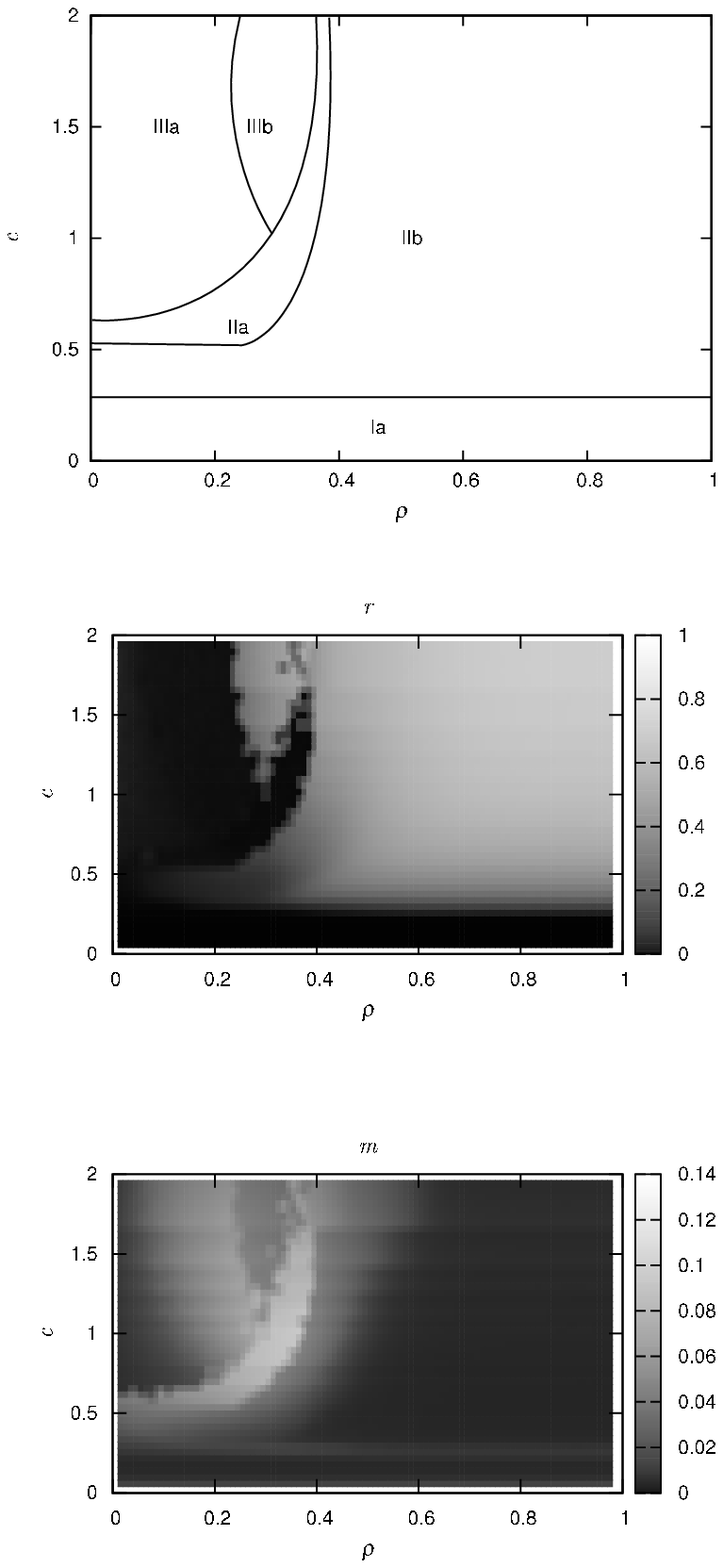}
\caption{\scriptsize{Top: Schematic of the different regimes in the $(\rho,c)$ plane of the parameter space. Lines were estimated by visually inspecting the dependence of the order parameter $r$ (middle) and the mean firing rate $m$ (bottom) on coupling strength $c$ and rewiring probability $\rho$. Initial conditions: $x_n(0) = 0$ for all neurons. The system was observed for 8000 time steps. I: random firing, II: self-sustaining patterns, III: cyclic waves. a: activity with only minor fluctuations, b: collective firing.} 
\label{smallworldOrderStartSynchronized}}
\end{figure}

In upper part of Fig. \ref{smallworldOrderStartSynchronized} we show a schematic of a partitioning of the $(\rho,c)$-plane into different regimes
derived from a visual inspection of the dependencies $m(\rho,c)$ and $r(\rho,c)$ (lower part). In addition to the regimes already described above (cf. Fig. \ref{meanOrderCoupling1.5}), we observe random firing (regime Ia) for coupling strengths $c < 0.28$. The transition to regime IIb is independent on $\rho$. This is to be expected since our estimation for the critical coupling strength $c_c$ does not depend on the network topology. For larger coupling strengths, the dynamics is influenced by the network topology, and particularly $\rho < 0.4$ allows for rich dynamical behavior. Counterintuitively,  we observe that an enhanced coupling strength can prevent the medium from periodic behavior (for example at $\rho=0.3$ and $c = 0.5$). Also an enhanced rewiring can prevent global oscillations (for example at $c = 1.5$ and $\rho = 0.3$).  Regime IIIb only exists for coupling strengths $c > 1$ and thus IIIa directly adjoins IIa for $c < 1$ as already observed for regular lattices.

As the network activity of both regular lattices and random networks is determined -- for large coupling strengths -- by the charging of the medium due to spontaneous firing, we here observe a low firing rate $m$. In the small-world regime, the medium is charged by the rewired connections of neurons on wavefronts, which leads to a high mean firing rate.

\begin{figure}
\includegraphics{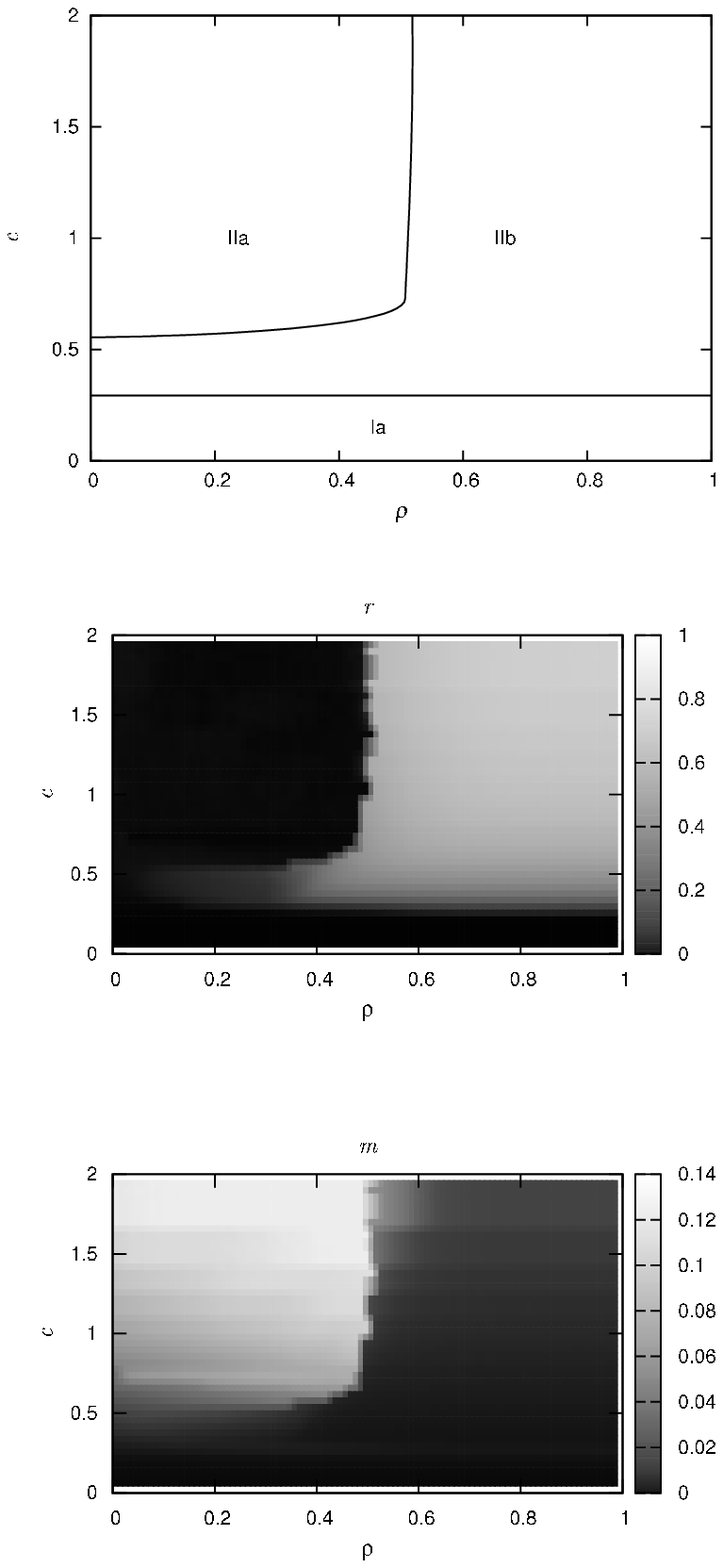}
\caption{\scriptsize{Same as Fig. \ref{smallworldOrderStartSynchronized} but for a spiral wave state as initial conditions.} \label{smallworldOrderStartDesynchronized}}
\end{figure}

Next, we study the stability of an initial spiral wave state depending on the coupling strength $c$ and rewiring probability $\rho$.
Since it is possible that the structure of a stable spiral changes with increasing $\rho$, initial conditions of a spiral wave taken from a regular lattice and applied to a small-world medium will possibly not lead to a stable spiral wave although such waves are still possible for this rewiring probability. For our simulation we started with a spiral wave for $\rho = 0$, which was created by imposing an open wave ending as initial conditions (see Fig. \ref{openWaveEnding}). After the emergence of a repetitive pattern and the usual measurement of $m$ and $r$, we imposed the membrane potential of each neuron as initial conditions for a medium with a one percent higher $\rho$. This procedure was iterated until $\rho = 1.0$. At some step the spiral wave is not stable anymore and begins pulsating with increasing amplitudes, which leads to global oscillations. This instability occurs when the spatial distance between two waves is large enough for neurons to be exited until threshold by long-range connections.  In Fig. \ref{smallworldOrderStartDesynchronized} we present our findings along this procedure. Surprisingly, we observe wave-dominated behavior for $0.4 < \rho < 0.5$ and large $c$ although in this regime global oscillations were stable when starting from $x_n(0) =0$ for all neurons. This is because the possibility of wave propagation depends on whether the medium shows global synchrony. For a wave traveling through a medium with global synchrony the mean excitation in the network $\bar{x}(t)$ increases on the way to collective firing. Wave propagation may only be possible at times where $\bar{x}(t)$ exceeds a certain level. In the case of a medium without global synchrony, however, the excitations from long-range connections are evenly distributed over time and allow for wave propagation as do local connections.  Waves are thus faster and are not moved backward during the collective firing as already mentioned above. Therefore, dynamical states of regime IIa and IIb can both be stable for $0.4 < \rho < 0.5$.

\section{Conclusions}
\label{sec:conclusion}
We investigated numerically the collective behavior of small-world networks of non-leaky IF neurons depending on the coupling strength $c$ and the number of random connections $\rho$. We considered a radius of influence $R$ for all nodes on the regular lattice underlying the small-world networks which, on the one hand, can be regarded a more realistic setting with respect to neuroanatomy and, on the other hand, led to a complicated dependence of the network dynamics on the coupling strength. Particularly, we observed turbulent-like patterns which can probably not be observed with nearest-neighbor couplings. Moreover, in the small-world regime, a radius of influence $R>1$ allowed for a fast and reproducible formation of spiral waves even without strong noisy inputs. Given our setup we observed different regimes in the ($\rho,c$)-plane which were characterized by different dynamical behaviors of the network. We observed local patterns such as cyclic waves, spiral waves, and turbulent-like patterns. For certain network parameters and depending on the set of initial conditions these patterns interfered with global collective firing of the neurons. Moreover, we observed in the same network different dynamics depending on the set of initial conditions only. Our observations indicate that both strength and topology of connections play an important role in determining the spatiotemporal dynamics of complex networks such as the brain during both physiological and pathophysiological conditions as can be observed e.g. in epilepsy. For the latter our findings are in line with those obtained from other modeling approaches \cite{Netoff2004,Percha2005,Dyhrfjeld-Johnsen2007,Feldt2007,Weihberger2007,Morgan2008} 
as well as with findings obtained from in vivo studies \cite{Ponten2007,Schindler2008a,Kramer2008} and emphasize the need for more experimental studies on functional and structural connectivity in real neural tissue. Progress along this line can be expected from recent methodological developments \cite{Buzsaki2004a,Michel2004,Sternickel2006} that allow one to study neural network activity at high spatial and temporal resolution.

\section*{Acknowledgements}
We are grateful to Stephan Bialonski, Anton Chernihovskyi and Marie-Therese Horstmann for helpful comments on earlier versions of the manuscript. This work was supported by the Deutsche Forschungsgemeinschaft (LE 660/4-1).


\end{document}